# Rate-dependent Stick-slips in Steady Shearing:

# Signature of Transition between Granular Solid and Fluid

Jih-Chiang (JC) Tsai [1*], Guan-Hao Huang [1], and Cheng-En Tsai [1,2]

[1] Institute of Physics, Academia Sinica, Taipei, Taiwan

[2] Department of Physics, Nat'l Central University, Chung-Li, Taiwan

Correspondence [(*)]: jctsai@phys.sinica.edu.tw

**Abstract** – Despite extensive studies on either smooth granular-fluid flow or the solid-like deformation at the slow limit, the change between these two extremes remains largely unexplored. By systematically investigating the fluctuations of tightly packed grains under steady shearing, we identify a transition zone with prominent stick-slip avalanches. We establish a state diagram, and propose a new dimensionless shear rate based on the speed dependence of inter-particle friction and particle size. With fluid-immersed particles confined in a fixed volume and forced to "flow" at viscous numbers $J$ decades below reported values, we answer how a granular system can transition to the regime sustained by solid-to-solid friction that goes beyond existing paradigms based on suspension rheology.

Granular-fluid mixtures, ranging from natural phenomena such as quicksand, debris flows, landslides, to grains in industrial mixers/silos, are excellent examples exhibiting solid-fluid duality. On one hand, smooth flows as a dense suspension have been successfully described by the paradigm of "$J$-rheology"[1] with viscous numbers $J$ down to the order of $10^{-5}$. On the other hand, flows of tightly packed grains have been treated by a separate tradition with elastoplastic models and theories[2]. Understanding the transition between these two extreme remains a profound challenge, not only in complexity of partial fluidizations[3–5], but also due to the lack of experimental observations on such transitions[6]. Two recent narratives since the initial proposal of the jamming phase diagram[7] have been that the "jamming point" depends on inter-particle friction[8], and that *friction* makes granular systems go beyond "ideal jamming"(by isotropic compression) and exhibit fragile states where the role of shearing becomes explicit[9–12]. Several experiments[13–15] and numerical studies[16] have demonstrated that friction plays a decisive role in creating dramatic change of behaviors.

Numerous prior works focus their attention on either rapid flows of suspension[13-15] or "quasi-static" systems[9,12], leaving the transition in-between largely unexplored. In reality, "friction coefficient" can vary substantially with the driving rate, and speed-dependent frictions have long been considered the root of instabilities leading to catastrophic consequences like earthquakes[18–20]. However, despite the on-going debates over the mechanism behind the rate-dependent jamming in rapid flows, also known as discontinuous shear thickening[13-17,27,28], roles of the speed dependence of friction on slow flows of tightly packed grains such as quicksand, steady flows in an hourglass, and their regime transition appear missing from existing discussions[6]. Recent studies have reported hysteresis and non-monotonic flow curves in experiments with frictional particles[21,22]. But without systematically looking into fluctuations, the *dynamics* behind the solid-fluid transition remains elusive.

In this Letter, we begin with showing a "dangerous zone" of driving rates that exhibits prominent stick-slip avalanches for packed grains sheared in a fixed volume. Such transition zone bridges the gap between the solid-like regime in the "slow" extreme and smooth flows on the faster side. Its occurrence is linked to the speed-dependent friction between particles, which we determine by direct measurements. This is interpreted as the consequence that low-friction contacts get a chance to percolate through a network of load-bearing particles and induce repetitive landslides. We establish a State Diagram that connects different flow regimes, and propose a new dimensionless shear rate based on the inter-particle tribology. Our work shows how the existing paradigm of $J$-rheology experiences its transition, as we create shear flows that are three decades below reported values of $J$ and reach the regime of granular solids.

*Setup and overview of transitions* --- Shown as in Fig.1(a), our particles fill the space between two cones. The sidewall is formed by a stack of free-sliding acrylic rings with an inner diameter $2R$=22cm. Two cones are geometrically roughed at the scale of the particle with diameter $d \approx$ 9mm. The upper cone is set at a fixed height, with the mechanism specially reinforced to provide a smooth rotation at an angular resolution of $2\pi \cdot 10^{-4}$ with a wide range of angular speeds $\Omega$. The base of the system is supported by the arrangement of six independent force sensors, in order to determine the total forces and moments that keep the base stationary. The gap between the rotating boundary and the sidewall is 2mm such that all particles cannot escape. The nominal volume fraction is defined as $\phi = N \, v_1 \, / V_{access}$ in which $N$=O(1000) is the number of particles. $V_{access}$ represents the total volume accessible by the particles, and $v_1$ stands for the volume for the single particle (which has been determined by Archimedes' method and checked for the mono-dispersity). Simultaneously, we take internal images of particles via the bottom, when the interstitial space is filled with aqueous solution of 60% glycerol such that the refraction index is matched to the particles[23,24]. We illuminate the system at its mid-height, with a horizontal laser sheet at the wavelength of 532nm going through the 1mm gap between the tips of two cones. Image contrasts are generated by dying either particles or the fluid with Rhodamine. Data reported in this paper are based on spherical particles of polydimethylsiloxane (PDMS) elastomer that we create by molding, with the standard curing agent provided by the manufacturer[25]. We have verified that particles exhibit a Hertzian response to compression with a Young's modulus around 1.5MPa, which is *soft* relative to the rigidity of our driving mechanism --- see Appendix-A and –C in SI[31] for further details.

To demonstrate the continuous change of behaviors, Fig.1(b) shows the time histogram of instantaneous torque $\tau_z$ (presented as an effective stress), for experiments at a wide range of driving rates [rps] sharing the same volume fraction. At either the fast or the slow end, the histogram presents a bell shape. Note that the slower runs are at the state of *higher* stress, in startling contrast with reported studies on most rheometric flows[26–28]. Interestingly, between the two extremes, histograms are highly asymmetrical (as shown by the parameter $b$). The case at 0.005rps is labelled as state-$T$. The extreme case with a relatively high (or low) stress is labelled as state-$\beta$ (or $\alpha$-) to facilitate subsequent discussions. Time series of $\tau_z$ for these three states are compared in Fig.1(c).

One might also time-average the $\tau_z$, define a mean stress $\sigma_S$, and plot it against the shear rate $\dot{\gamma} \equiv \Omega$ /2tan $\xi$: Shown as in Fig.1(d), the behaviors mimic the non-monotonic "flow curves" as those reported by Ref.[22]. Note that Fig.1(d) cover data from experiments at a wider range of driving rates than those in Fig.1(b), plus two additional runs with interstitial fluids at higher viscosities (by increasing the concentration of glycerol). The mean stress shows not only an interval of negative slope over $\dot{\gamma}$, but also an uprising across its minimum. As a rough indicator for the occurrence of stick-slips, we compute the normalized counts of large drops that we refer as LD, that is defined as a sudden drop of torque larger than twice the root-mean-square of the total fluctuation --see Appendix-B in SI[31]). Results are plotted in Fig.1(e) against the viscous number[1] $J \equiv \eta \, \dot{\gamma} \, / \sigma_N$ which incorporates our normal-stress measurements $\sigma_N \equiv$ time-averaged $F_z$ /$\pi R^2$. The data show a consistent rise-and-fall over the change of $J$.

*Stick-slip avalanches & the slow extreme* --- The change from state-$T$ to state-$\beta$ deserves special attention. In, Fig.2, we show the cumulative distributions of stress drops as functions of a variable threshold $\Delta_{min}$: For each driving rate, we count the number of sudden drops in torque (-$\Delta \tau_z$) that are larger than $\Delta_{min}$, normalized by the total strain, and plot the normalized counts against $\Delta_{min}$. The results characterize the change of behaviors in a quantitative manner.

For **state-$T$**, we use the inset to highlight that the *small* avalanches resembles a power-law distribution up to about 40% of the mean torque $<\tau_z>$. This can be translated into a probability density $\sim |\Delta \tau_z|^{-1.7}$. It is worth noting that observations of power laws can be traced back to centuries of earthquake observations, known as Gutenburg-Richer Law showing a power-law distribution on released energy $\sim E^{-\tau_0}$ extending for several decades: Reported values of $\tau_0$ ranges from 1.8 to 2.2 depending on geography, and are reproduced by numerical models in different degree of conservation[34]. In terms of stress release, mean-field-theory prediction has a long tradition back to half a century ago and $\tau_0=3/2$ is commonly seen[35].

On the other hand, *larger* avalanches deviate strongly from power law and shows a stronger-than-exponential truncation --- see dashed line and the description in caption. The suggests that large slips represents different physics from small ones. (Similar distinctions have been discussed over classic studies on sand-piles in terms of system-spanning flushing versus localized failures[36]). Quantitatively, the statistics show that stress drops larger than 1kPa (that corresponds to $\sim 0.3 < \tau_z >$) can occur several times within a shear strain of unity. Readers are encouraged to watch online movies[31] of differential images that reveal sudden, system-spanning displacements of particles upon these stress drops ---- Fig.1(f) shows four snapshots around one such event. Also available online are animations based on the multi-component force measurements: For **state-$T$**, the animated $\tau_z$-$F_z$ plot reveals multiple "loading curves" that are often terminated by a slip event. Meanwhile, the scattered plot of the net force $F_y$-$F_x$ show "clusters" during the stress build-ups and sudden "jumps" in-between.

For **state-$\beta$**, one might see the granular pack as mostly in the *stick* mode, at a relatively high level of stress. The system "flows" in the form of small, isolated avalanches --- see movies on the online SI[31] for their spatial sizes and vivid demonstrations of "dynamical heterogeneities"[29]. Statistically, Fig.2 shows that the stress drop above 1kPa would be *extremely rare*: it takes a shear strain $\sim 10^2$ (corresponding to about three days in our laboratory experiments) to accumulate one. Such rareness indicates the lack of opportunities to "reset" the grain configurations and explains why state-$\beta$ displays a higher mean stress than that of state-$T$. Recalling the time distribution of $\tau_z$ (Fig.1b), it is also important to point out that even though state-$\beta$ displays a bell shape resembling that of state-$\alpha$, the two states are dominated by qualitatively different physics. As a result, at precisely the same volume fraction, the time-averaged $\sigma_S$ is not only higher at state-$\beta$ but also shows a plateau, as seen from Fig.1(d) and our further elaboration below.

*Connection with tribology and a State Diagram for solid-fluid transition* – These rate-dependent behaviors can be consistently explained by first considering two facts. (1) Illustrated as Fig.3(a), imposing a shear rate $\dot{\gamma}$ to a densely packed system creates a range of relative speeds among particles. In a system that is strongly damped, the anticipated distribution should be around the most probable speed $\sim d\,\dot{\gamma}$. (2) By independent measurements using the same fluid and PDMS elastomer with a radius of curvature $R=d/2$ to mimic our main experiments, we characterize the speed-dependent friction between surfaces – see Fig.3(b). The results show a characteristic speed $V_c$ beyond which the tangential force decreases dramatically. Such effects are consistent with the transition from the regime of solid-solid contact to that of mixed lubrication, that is commonly cited as part of the Stribeck diagram in tribology[14,30]. We have also confirmed that $V_c$ is inversely proportional to the fluid viscosity, in consistence with prior literatures [30].

We are now in a position to summarize our observations in a *State Diagram*, shown as Fig.3(c). The main plot contains multiple datasets with different combinations of volume fractions and fluid viscosity. We display the time-averaged shear stress $\sigma_S$ as its vertical coordinate, over a horizontal plane spanned by the mean normal stress $\sigma_N$ and a dimensionless shear rate $d\dot{\gamma}/V_c$. Even without the full knowledge of the statistical distribution of inter-particle speeds, one can anticipate that for state-$T$ to exhibit stick-slip behaviors, thee conditions are necessary: (1) Some of the particles must be in solid-solid contact (with a relative speed below $V_c$.) to build up a highly stressed network. (2) Meanwhile, the spectrum of inter-particle speeds should go across $V_c$, such that a fraction of contacts are in the lubricated state, behaving as the "weak spots". These weak spots migrate as the macroscopic shearing goes on and are likely to trigger a cascade of avalanche once they get a chance to percolate. (3) The number of solid-to-solid versus lubricated contacts should be comparable for the build-up and avalanche to reoccur with substantial amplitudes. Indeed, our data reveal that stick-slip is prominent when $0.1 < \left(\frac{d\dot{\gamma}}{V_c}\right) < 1$. On one hand, when $\frac{d\dot{\gamma}}{V_c} \ll 1$, one could imagine that the number density of weak spots become too low, such that the system can only create sporadic, localized slips. In the other extreme, when $\frac{d\dot{\gamma}}{V_c} > 1$, the majority of contacts are lubricated such that stress build-up is prohibited such that the stress stays low. However one can anticipate a non-monotonic change as the viscous drag becomes dominant at higher shear rates and, far above the current scope of our experiments, shear thickening effects[13–17,27,28] might occur and demand descriptions by inertial numbers[37] instead.

Understandably, a clear separation between the regime dominated by solid-solid contacts and that by lubricated sliding demands a high volume fraction that warrants a network of load-bearing particles --- in our case with $\phi > RLP$ [8]: We have verified by further data (Appendix-E in SI[31]) that, as $\phi$ goes below $RLP$, (a) the stick-slips behaviors separating the two regimes mentioned above become undetectable, and that (b) at $\phi = 0.48$, $\sigma_S$ ($\dot{\gamma}$) becomes monotonic and bears the appearance of the conventional "flow curves", resembling what we have demonstrated previously with hydrogel particles in the same apparatus[32]. On Fig.3(c), we use a diagram on the right to illustrate that the stick-slips (the shaded area) disappear such that two regimes merge, once the system becomes a true suspension --- with a low normal stress $\sigma_N$ due to insufficient volume fractions which prevent the formation of a highly stressed packing. We find that the merging occurs with a viscous number $J$ = $10^{-6}$ and $d\dot{\gamma}/V_c \sim 0.3$. This observation is consistent with one recent work with a rotating drum[21] that reports a critical value $J_c \sim 10^{-6}$, below which smooth flows cease to exist. In fact, our experiments with grains packed in a fixed volume has extended the observations down to $J \sim 10^{-9}$ that is three decades below reported values in most stress-controlled experiments reviewed by Ref. [1]. The smallness of the new dimensionless parameter $d\dot{\gamma}/V_c$ provides a definition of "quasi-static solids".

Interestingly, our discussions on how a dense granular pack ceases to be fluidic and enters the solid-like regime as the driving rate decreases also shares many vocabularies with the BDH model proposed in 2011[33] ---- see Appendix-F in SI[31] for a side-by-side comparison in the similarity and key differences.

*Concluding Remarks*--- By recognizing the speed-dependent friction between particles, we establish *a new dimensionless shear rate $d\dot{\gamma}/V_c$* for describing the solid-like behaviors that are not covered in existing paradigms based on the rheology of dense suspension. We find that (1) at high packing fractions, our data demonstrate a "dangerous zone" of shear rates exhibiting prominent stick-slips, that provide an unambiguous signature to divide the solid-like regime (dominated by quasi-static contacts) and the fluidic regime (dominated by lubricated sliding); that (2) evidence from boundary force fluctuations combined with internal imaging reveals that such solid-fluid transition takes place in a continuous yet distinctive manner; and that (3) the transition is controlled by two parameters: driving rate and $\phi$ (or the stress $\sigma_N$), not just $\phi$ alone.

Our interpretations distinguish from existing paradigms in three aspects. (A) We emphasize that the tangential force between contacting particles is sensitive to the relative speeds. Taking the speed distribution into account, one might see the system as having a mixture of high- versus low-friction contacts. To the best of authors' knowledge, instabilities induced by *distributed* frictional coefficients have not yet been considered in most existing theories or models. In our opinion, this is essential in understanding systems in which velocity weakening is known to exist such as in earthquakes or quicksand. (B) The speed-dependent friction makes the packing fraction $\phi$ no longer the leading factor for the solid-fluid transition. (C) Features of the particle contacts are now considered in the dimensionless parameter we propose, unlike how they appear missing from previous paradigms such as in viscous number $J$. The regime changes are summarized in the state diagram we provide. We hope to inspire further interpretations from various points of view, such as a generalization from theories based on the percolations[10] or connections with elastoplastic models as shear rates are varied[38].

--------

JCT acknowledges fruitful exchanges with X. Cheng, J. Dijksman, 梁鈞泰, 陳志強, 黃仲仁, 施宏燕 and Nigel Goldenfeld on the manuscript, and supports of IoP especially from the Precision Machine Shop.

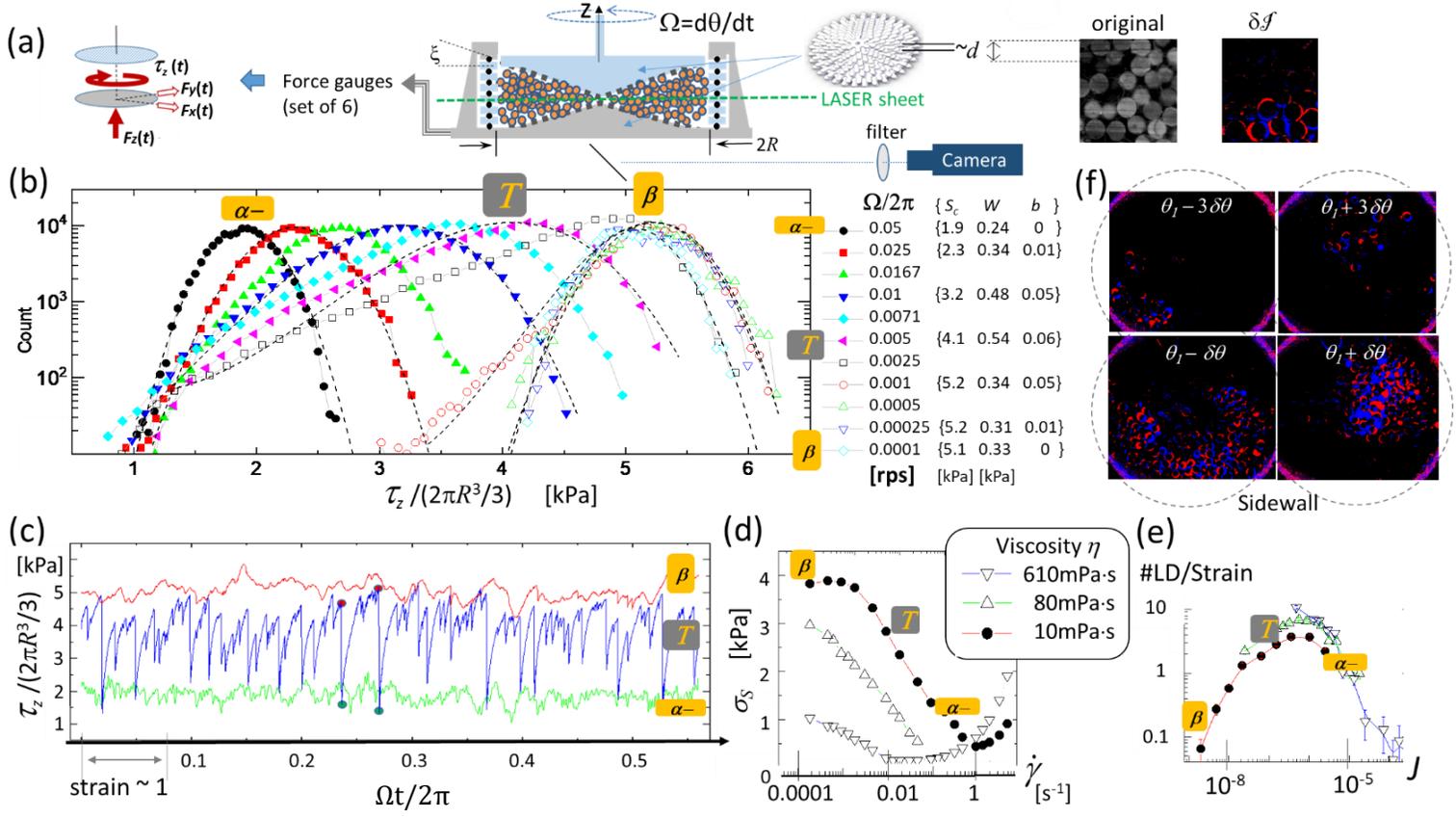

**FIG. 1 (a)** Schematics for the setup and measurements, from left to right: the time-dependent torque and three net forces as determined from sensors around the base; cross-sectional view of the main setup (with a stack of acrylic rings alternating with ball bearings to minimize resistance along $\theta$); 3D illustration of the roughened cones; close-up of a typical fluorescent image; and the differential image $\delta\mathcal{I}$ computed from two frames with $\delta\theta=8\pi\cdot10^{-4}$. **(b)** Statistical distributions of torque $\tau_z$ measured over time, at different rotation rates $\Omega/2\pi$ [in revolutions per second, rps] denoted by symbols. Dashed lines show curves $\sim \exp\{-0.5\left(\frac{S-S_c}{W}\right)^2 - b\left(\frac{S-S_c}{W}\right)^3\}$, where $S \equiv \tau_z/(2\pi R^3/3)$ with empirical fit parameters indicated on the graph. $\phi=0.60$. **(c)** Time series of $S$ for three states in (b), all plotted against the angular displacement, which can also be converted to effective *strain* $\equiv \Omega t/2\tan\xi$. The high-frequency noises are smoothed by 40ms. **(d)** Time-averaged stress $\sigma_S$, plotted against the shear rate $\dot{\gamma} \equiv \Omega/2\tan\xi =13.7(\Omega/2\pi)$. **(e)** Counts of LD *per unit strain*, plotted against the viscous number $J$, as defined in main texts. Data in (d) and (e) involve experiments with three different viscosity $\eta$ in the fluid, at the same $\phi=0.56$. **(f)** Sequence of differential images $\delta\mathcal{I}$ with $\delta\theta=2\pi\cdot10^{-4}$, accompanying one of the large stress jumps in state-$T$. Pixels in positive (negative) values are displayed in blue (red). The boundary movement $\theta(t)$ are indicated in reference to a fixed $\theta_1$.

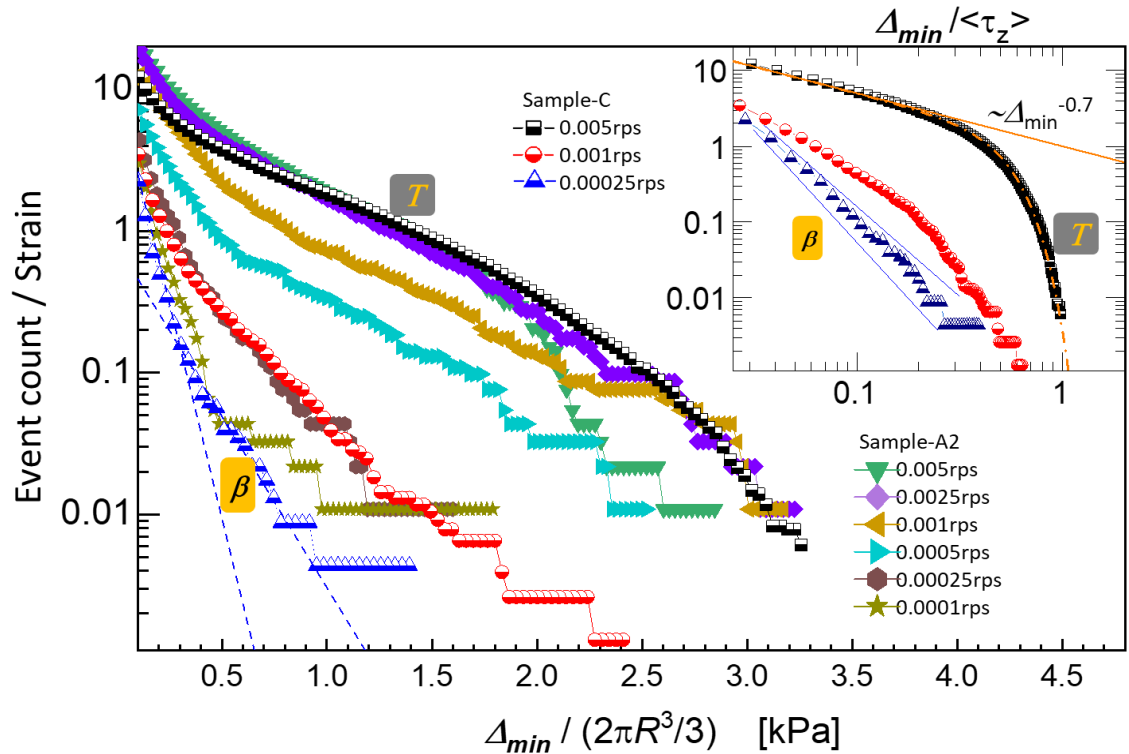

**FIG.2** Cumulative distribution of torque (stress) drops with $-\Delta\tau_z > \Delta_{min}$, for experiments at different driving rates [rps]. Event counts are normalized by the strain accumulated. Data cover experiments using two batches of particles (A2 and C) in the same fluid. $\phi=0.56$. In the inset, $\Delta_{min}$, is normalized by the mean $<\tau_z>$ for each driving rate, while logarithmic slopes of $-0.7$, $-2.5$, and $-3$ are indicated by straight lines and a truncated power law $\sim \left(\frac{\Delta_{min}}{<\tau_z>}\right)^{-0.7} \exp\left\{-\left(\frac{\Delta_{min}}{<\tau_z>}/0.54\right)^{2.8}\right\}$ by the dash-dot curve--- see Appendix-D in SI[31] for further analyses. In the main plot, two dotted lines illustrate exponential decays with the characteristic stress being 0.2kPa and 0.4kPa, respectively. Although statistics from the two batches of particles do not coincide exactly in terms of the corresponding rps (due to different histories of immersion ----see Appendix-C3 in SI), the general trend over the wide range of driving rates is consistent.

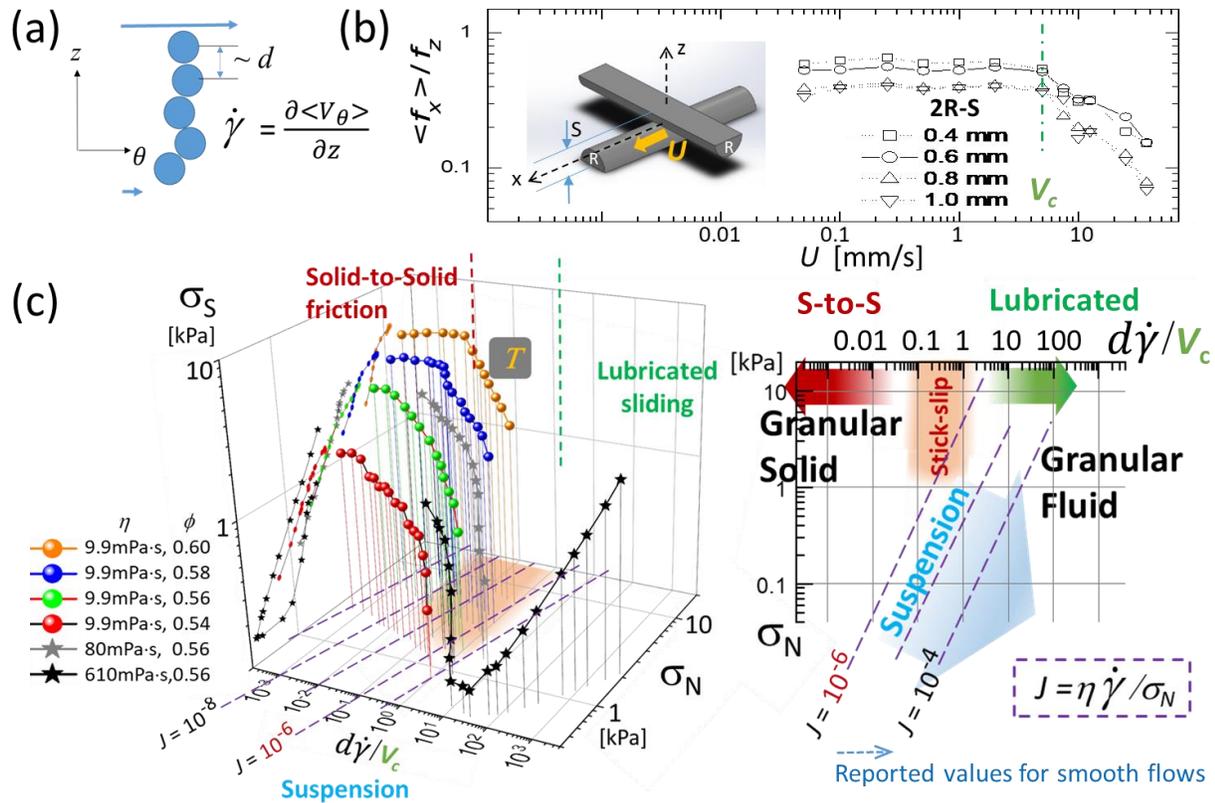

**FIG.3 (a)** Illustration of the motion of densely packed particles under an imposed shear rate $\dot{\gamma}$ ; **(b)** Schematics and results of measurements on the average tangential force $f_x$, normalized by the normal force $f_z$, for a localized contact between PDMS surfaces with a radius of curvature $R$. Data are plotted as functions of the sliding speed, for experiments at multiple pressing depths (controlled by the distance $S$ between the base planes) up to about $0.1d$. Data reveals a critical velocity $V_c$ that appears insensitive to the pressing depth $2R$-$S$. **(c)** State diagram summarizing data from experiments with different values of $\phi$ and $\eta$. On the main graph, the vertical coordinate shows the time-averaged shear stress $\sigma_S$, plotted against the dimensionless shear rate $d\dot{\gamma}/V_c$ and the normal stress $\sigma_N$. On the right, we display a schematic projection onto the $\sigma_N$-$d\dot{\gamma}/V_c$ plane, with contours of constant viscous number $J$ represented by dashed lines indicating $10^{-4}$, $10^{-6}$, and in turn to values as small as $10^{-8}$ back on the main graph. Reported values of $J$ are based on Ref.[1].